\documentclass[a4paper,fleqn,usenatbib]{mnras}

\usepackage{mathptmx}

\usepackage[T1]{fontenc}
\usepackage{ae,aecompl}

\usepackage{graphicx}	
\usepackage{amsmath}	
\usepackage{amssymb}	

\title[Intermittency spectra of current helicity]{Intermittency spectra of current helicity in solar active regions}

\author[A. S. Kutsenko et al.]{
A. S. Kutsenko,$^{1}$\thanks{E-mail: alex.s.kutsenko@gmail.com (ASK)}
V. I. Abramenko,$^{1}$
K. M. Kuzanyan,$^{2}$
Haiqing Xu$^{3}$
and Hongqi Zhang$^{3}$
\\
$^{1}$Crimean Astrophysical Observatory, Nauchny, Crimea 298409, Russia\\
$^{2}$IZMIRAN, Troitsk, Moscow 108840, Russia\\
$^{3}$National Astronomical Observatories, Chinese Academy of Sciences, Beijing 100012, China
}

\date{Accepted XXX. Received YYY; in original form ZZZ}

\pubyear{2018}

\begin{document}
\label{firstpage}
\pagerange{\pageref{firstpage}--\pageref{lastpage}}
\maketitle

\begin{abstract}
We analyse the spatial distribution of current helicity in solar active regions. A comparison of current helicity maps derived from three different instruments (Helioseismic and Magnetc Imager on board the \textit{Solar Dynamics Observatory}, SDO/HMI, Spectro-Polarimeter on board the \textit{Hinode}, and Solar Magnetic Field Telescope at the Huairou Solar Observing Station, China, HSOS/SMFT) is carried out. The comparison showed an excellent correlation between the maps derived from the spaceborne instruments and moderate correlation between the maps derived from SDO/HMI and HSOS/SMFT vector magnetograms. The results suggest that the obtained maps characterize real spatial distribution of current helicity over an active region. To analyse intermittency of current helicity, we traditionally use the high-order structure functions and flatness function approach. The slope of a flatness function within some range of scales -- the flatness exponent -- is a measure of the degree of intermittency. SDO/HMI vector magnetograms for 3 ARs (NOAA 11158, 12494, and 12673) were used to calculate the flatness exponent time variations. All three ARs exhibited emergence of a new magnetic flux during the observational interval. The flatness exponent indicated the increase of intermittency 12--20 hours before the emergence of a new flux. We suppose that this behaviour can indicate subphotospheric fragmentation or distortion of the pre-existed electric current system by emerging magnetic flux.
\end{abstract}




\section{Introduction}
\label{sec:intro}
Kinetic and magnetic helicities play an important role in the mean-field dynamo theory \citep[e.g.][]{Parker1979, Charbonneau2005}. In the framework of the theory, velocity $\mathbfit{v} = \langle\mathbfit{v}\rangle + \mathbfit{v}'$ and magnetic $\mathbfit{B} = \langle\mathbfit{B}\rangle + \mathbfit{B}'$ fields are split up into mean ($\langle\mathbfit{v}\rangle$ and $\langle\mathbfit{B}\rangle$) and fluctuating ($\mathbfit{v}'$ and $\mathbfit{B}'$) parts. The induction equation for  the mean magnetic field, $\langle\mathbfit{B}\rangle$, can be written as \citep[e.g.][]{Seehafer1994}:
\begin{equation}
\frac{\partial \langle\mathbfit{B}\rangle}{\partial t} = \nabla\boldsymbol{\times}(\langle\mathbfit{v}\rangle \boldsymbol{\times} \langle\mathbfit{B}\rangle) + \eta \Delta \langle\mathbfit{B}\rangle + \nabla\boldsymbol{\times} \mathcal{E},
\label{eq0}
\end{equation}
where $\eta$ is the magnetic diffusivity constant and $\mathcal{E}$ is the mean electromotive force caused by the fluctuations:
\begin{equation}
 \mathcal{E} = \langle\mathbfit{v}' \boldsymbol{\times} \mathbfit{B}' \rangle.
\label{eq05}
\end{equation}
Under a set of assumptions, supposing that i) the spatial scales of the mean and fluctuating parts of fields can be separated and ii) the fluctuations are statistically homogeneous, a relationship must hold \citep{Seehafer1994}
\begin{equation}
\mathcal{E} = \alpha \langle\mathbfit{B}\rangle,
\label{eq1}
\end{equation}
with $\alpha$ denoting some scalar or pseudo-tensor. The electromotive force can be the only source of the mean-field magnetic energy increase \citep{Seehafer1994}. Expression \eqref{eq1} describes the so-called alpha effect that requires a non-vanishing kinetic helicity \citep{Seehafer1990}. It is assumed that the Coriolis effect can be responsible for the generation of helicity. Since the Coriolis effect produces helical motions of opposite direction in the northern and in the southern hemispheres, a certain helicity sign must prevail in the northern hemisphere and the opposite sign must prevail in the southern hemisphere. Unfortunately, neither kinetic nor magnetic helicity are available for direct observations. Nevertheless, one can adopt current helicity, \textit{H\textsubscript{C}}, as a proxy of magnetic helicity: current and magnetic helicities have the same sign and they increase simultaneously \citep{Seehafer1990}.

The hemispheric current helicity sign rule was established in the pioneering work by \cite{Seehafer1990}. He analysed the curvature of fibrils in the vicinity of 16 active regions (ARs) to choose the sign of the parameter \textit{$\alpha$}
\begin{equation}
\nabla\boldsymbol{\times}\mathbfit{B} =\alpha\mathbfit{B}
\label{eq2}
\end{equation}
calculated in a force-free field approximation. The author concluded that current helicity
\begin{equation}
H\textsubscript{C}=\mathbfit{B}\boldsymbol{\cdot}\nabla\boldsymbol{\times}\mathbfit{B} = \alpha\mathbfit{B}\textsuperscript{2}
\label{eq3}
\end{equation}
is predominantly negative in the northern and positive in the southern hemisphere. A series of works confirmed his results with larger statistical samples \citep{Pevtsov1994, Pevtsov1995, Abramenko1996, Bao1998, Hagino2004}. It was found in a number of works that the hemispheric sign rule can be violated during certain cycle phases and at certain latitudes \citep{Bao2000,  Hagino2005, Pevtsov2008, Zhang2010,Otsuji2015}. Nevertheless, about 70-80\% of ARs obey the hemispheric sign rule \citep[e.g.][]{ Liu2014}.

Majority of works regard current helicity as some integral characteristic of an AR and focuses on its sign \citep[e.g.][]{Seehafer1990}, averaged value over an AR \citep[e.g.][]{Zhang2010}, imbalance \citep{Abramenko1996}, etc. Analyses of the spatial distribution of current helicity over an AR are not so numerous \citep[e.g.][]{Pevtsov1994,Zhang2006, Hao2011, Otsuji2015}. Thus, \cite{Pevtsov1994} found that the current helicity elements have a characteristic size of order 10 Mm and characteristic lifetime of order 27 hours. In this paper, we focus on the spatial distribution of current helicity over ARs and, for the first time to our knowledge, on its intermittency.
	
\section{Data}

In this study, we used vector magnetic field obtained by three instruments. The first instrument is the space-based Helioseismic and Magnetic Imager on board the \textit{Solar Dynamics Observatory} \citep[SDO/HMI,][]{ Scherrer2012, Schou2012}. SDO/HMI is a filtergraph that observes the solar disk at the \ion{Fe}{i} 6173~\AA\ photosphere spectral line. The spatial resolution of the instrument is 1\arcsec\ with 0.5\arcsec\ pixel size \citep{Liu2012}. Continuous set of the HMI magnetic field data with a cadence of 720 s is available since May 2010. The vector magnetograms are publicly available as full-disk 4096$\times$4096 pixel maps. Alternatively, the data are provided in the form of the Space-Weather HMI Active Regions Patches \citep[SHARPs,][]{Bobra2014, Hoeksema2014}. The SHARP data contain vector magnetic field maps of ARs automatically identified by a sophisticated algorithm \citep{Turmon2010}. ARs are tracked for its entire lifetime as they pass the solar disk. The 180-degree ambiguity in the transverse component of the HMI magnetic field vector is resolved using a minimum energy method \citep{Metcalf1994, Leka2009}. In this study, we used definitive $hmi.sharp720s$ data series.

The second instrument we used is the space-based Solar Optical Telescope Spectro-Polaimeter \citep[SOT-SP,][]{Lites2013} on board the joint Japanese/US/UK space mission \textit{Hinode} \citep{Kosugi2007}. Hinode/SOT-SP provides measurements of all four Stokes parameters measured at the \ion{Fe}{i} 6302~\AA\ lines. Only a one-dimensional slice of the solar surface is recorded at a time so the map of a certain region is obtained as a set of consecutive slices. Four operational modes are available. In Normal Mode the instrument can obtain a 1024$\times$1024 pixel map of the vector magnetic field of ARs with 0.15\arcsec$\times$0.16\arcsec\ pixel size during approximately 90 minutes. In Fast Map Mode the binning of the neighbouring pixels is performed resulting in 0.30\arcsec$\times$0.32\arcsec\ spatial samplings. It takes 30 minutes to acquire the vector magnetic field map of an AR with 151\arcsec$\times$162\arcsec\ size \citep{Lites2013}. We used Hinode/SOT-SP Level2 data that are outputs from MERLIN spectral line inversion code \citep{Lites2007}. Each Level2 dataset is stored in a FITS file and contains 42 extensions. The first three extensions are maps of magnetic field strength, inclination, and azimuth angles. To resolve the 180-degree ambiguity in the Hinode/SOT-SP vector magnetic field data we applied the new disambiguation code developed by \cite{Rudenko2014}.

The third instrument we used is the ground-based Solar Magnetic Field Telescope (SMFT) at the Huairou Solar Observing Station (HSOS), National Astronomical Observatories of China. The SMFT is equipped with a birefringent filter for wavelength selection and KDP crystals to modulate polarization signals. The \ion{Fe}{i} 5324.19~\AA\ spectral line is used at the HSOS vector magnetograph. A vector magnetogram is built using four narrow-band (0.125 \AA) filtergrams of Stokes $I$, $Q$, $U$, and $V$ parameters. The centre wavelength of the filter can be shifted and is normally at -0.075 \AA\ from the line centre for the measurements of longitudinal magnetic field and at the line centre for the transversal magnetic fields \citep{Ai1986}. The instrument has been observing vector magnetic fields for nearly 30 years. Three different CCD cameras have been used: prior to December 2001, observations were taken with a 512$\times$512 pixel CCD with the effective field of view of 5.23\arcmin$\times$3.63\arcmin. Between 2002 and 2008, a detector was replaced to 640$\times$480 pixel CCD with the effective field of view of 3.75\arcmin$\times$2.81\arcmin. After 2009, a 992$\times$992 pixel CCD is used, the effective field of view is 4\arcmin$\times$3.5\arcmin, resulting in a pixel size of about 0.30\arcsec$\times$0.29\arcsec. The weak-field approximation is used to reconstruct the vector magnetogram, the relationship between the magnetic field and the Stokes parameters $I$, $Q$, $U$, and $V$ is as following:
\begin{equation}
\begin{aligned}
B\textsubscript{l}=C\textsubscript{l}(V/I),\\
B\textsubscript{t}=C\textsubscript{t}\lbrack(Q/I)\textsuperscript{2}+(U/I)\textsuperscript{2}\rbrack,\\
\phi=0.5\tan\textsuperscript{-1}(U/Q)
\end{aligned}
\label{eq4}
\end{equation}
where $C\textsubscript{l}$ and $C\textsubscript{t}$ are the calibration coefficients for the longitudinal $B\textsubscript{l}$ and transverse $B\textsubscript{t}$ magnetic field, respectively, $\phi$ is the azimuth angle. For more detailed description of SMFT calibration, see e.g. \cite{Wang1996, Su2004, Bai2014}.

\section{Methods}
\subsection{Current helicity}
\begin{figure*}
	\includegraphics[width = \linewidth]{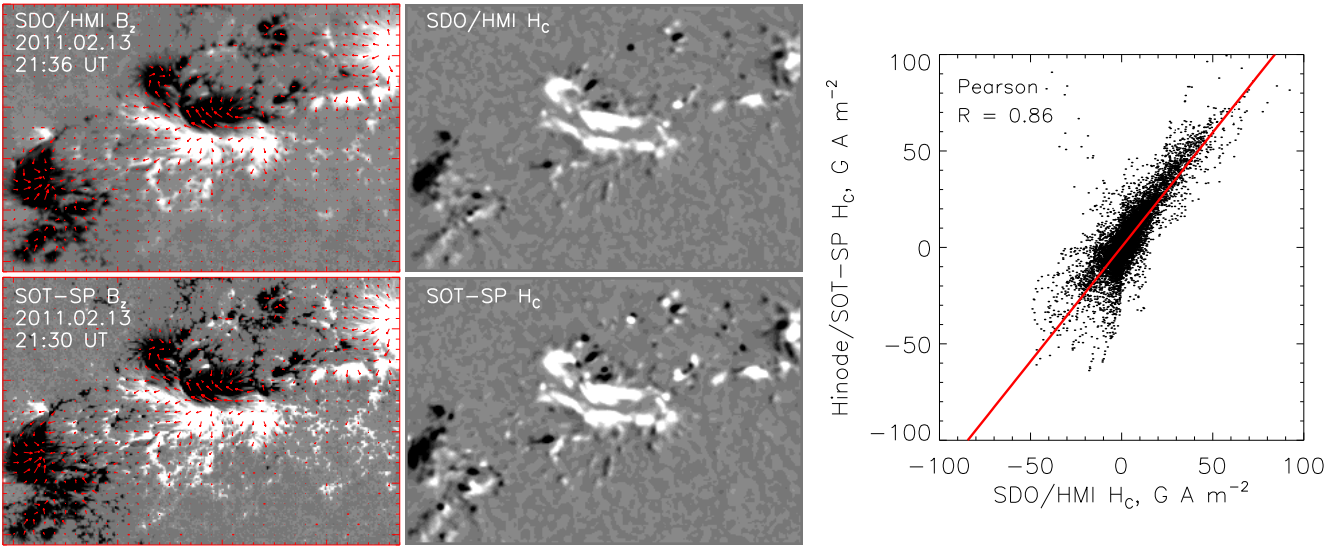}
	\caption{Left: Longitudinal magnetograms of AR NOAA 11158 acquired by SDO/HMI and Hinode/SOT-SP. The magnetograms were taken on 2011 February 13. The FOV is 150\arcsec$\times$100\arcsec. The magnetograms are scaled from -1000 Mx cm\textsuperscript{-2} (black) to 1000 Mx cm\textsuperscript{-2}. Red arrows indicates the direction of the transversal magnetic field. Middle: Current helicity maps of the AR derived from the magnetograms by equation \eqref{eq5}. The maps are scaled from -100 G A m\textsuperscript{-2} (black) to 100 G A m\textsuperscript{-2}. Right: A comparison of the current helicity maps. Thick red line shows the best linear fit to the distribution.}
	\label{fig1}
\end{figure*}

Using the full-vector magnetic field measurements at a single height in the photosphere, we can calculate only a z-related part of current helicity:
\begin{equation}
H\textsubscript{C\textsubscript{z}}=B\textsubscript{z}j\textsubscript{z}
\label{eq5}
\end{equation}
where $j\textsubscript{z}$ is the z-part of electric current density
\begin{equation}
j\textsubscript{z}=\lbrack\nabla\boldsymbol{\times}\mathbfit{B}\rbrack\textsubscript{z}
\label{eq6}
\end{equation}
Throughout this article, we will refer to $H\textsubscript{C\textsubscript{z}}$ as current helicity, $H\textsubscript{C}$. Using the Stokes' theorem and following \cite{Abramenko1996}, we can rewrite \eqref{eq6} in the following form:
\begin{equation}
j\textsubscript{z}=\oint_L\mathbfit{B}\textsubscript{t}\textrm{d}\mathbfit{r}
\label{eq7}
\end{equation}
where $\mathbfit{B}\textsubscript{t}$ is the vector of the transverse magnetic field. The integration was performed by Simpson's formula over the contour $L$ enclosed an area of $n \times n$ pixels around the central point where the z-part of the electric current $j\textsubscript{z}$ was determined. Electric currents obtained by equation \eqref{eq7} are in a good agreement with that obtained by a traditional differential formula with subsequent smoothing \citep{Abramenko1996}. In addition, the current helicity map of AR NOAA 11158 in Figure~\ref{fig1} is in an excellent visual agreement with almost co-temporary map in fig.~1 in \cite{Zhang2016}.

\subsection{Intermittency spectra}

Within the next sections, we analyse the intermittency of current helicity and magnetic fields distributions over ARs. We use the high-order structure function approach \citep[e.g.][]{Frisch1995, Abramenko2002, Abramenko2005}.

Structure functions were introduced by \cite{Kolmogorov1941, Kolmogorov1991} and they represent the statistical moments of the \textit{q}-powers of the increment of any two-dimensional (in our case) field $u(\mathbfit{r})$:
\begin{equation}
S\textsubscript{q}(\mathbfit{r})=\langle\lvert u(\mathbfit{x}+\mathbfit{r})-u(\mathbfit{x}) \rvert\textsuperscript{q}\rangle
\label{eq8}
\end{equation}
Here $\mathbfit{x}$ is each point of the analysed field and $\mathbfit{r}$ is the separation vector between the points used to measure the increment. The order of a statistical moment \textit{q} can take real values. The angular brackets in equation \eqref{eq8} denote averaging over the whole field map. More detailed description of the calculation procedure and of the $S\textsubscript{q}(\mathbfit{r})$ function properties can be found in \cite{Abramenko2002}.

At the next step we are to determine the scale range where $S\textsubscript{q}(\mathbfit{r})$ function is linear and the analysed field is intermittent. To do this, we calculate the hyper-flatness function determined as the ratio of the sixth statistical moment to the cube of the second statistical moment \citep{Abramenko2005}:
\begin{equation}
F(\mathbfit{r})=S\textsubscript{6}(\mathbfit{r})/(S\textsubscript{2}(\mathbfit{r}))\textsuperscript{3}\sim\mathbfit{r}\textsuperscript{-$\kappa$}
\label{eq9}
\end{equation}

For an intermittent field, $F(\mathbfit{r})$ changes as a power law of the scale $\mathbfit{r}$. In a double logarithmic plot of $F(\mathbfit{r})$ versus $\mathbfit{r}$, the slope $\kappa$ of the hyper-flatness function, determined within some scale range $\Delta\mathbfit{r}$ where $F(\mathbfit{r})$ is linear, characterizes the intermittency of the field: the higher the slope the higher the complexity of the spatial structure of the analyzed field \citep{Abramenko2005}. For simplicity, we will refer to $F(\mathbfit{r})$ and $\kappa$ as the flatness function and the flatness exponent, respectively, while the subscript will denote the analysed two-dimensional field (for example, $\kappa_{H_C}$ for the flatness exponent of the current helicity map).

Intermittency implies a tendency of a physical entity to concentrate into small-scale features of high intensity surrounded by extended areas of less intense fluctuations and  manifests itself via burst-like behaviour in temporal and spatial domains. Large fluctuations in an intermittent process are not as rare as in the Gaussian process, and they contribute significantly into the statistical moments, that leads to multifractality \citep[see, e.g.,][for more details]{Frisch1995}. Thus, by analysing multifractality of a structure, we also explore its intermittency. Generally speaking, intermittency and multifractality are two different terms of the same phenomenon. Historically, the term 'intermittency' is usually applied to time series analysis, whereas the term multifractality is used for spatial objects. In \citet{Abramenko2010} it was shown that the classical spectrum of multifractality $f(\alpha)$ versus the singularity exponent $\alpha$ can be derived from structure functions of a set of statistical moments, $q$. One of the reasons to use the aforementioned flatness function technique instead of the classical method is the fact that the flatness functions clearly reveal the scale interval, $\Delta\mathbfit{r}$, where  intermittency/multifractality holds (which is not the case of sophisticated singularity exponent), and besides that, the slope of the flatness function allows us a numerical estimate of the degree of intermittency (which is also a challenge when using the $f(\alpha)$ method).

\section{Comparison of current helicity maps from different instruments}

As it was mentioned above, majority of researchers regard current helicity as some integral characteristic of an AR, i.e. they examine the current helicity imbalance over a map, predominant sign, mean value, etc \citep{Seehafer1990, Zhang2010, Abramenko1996}. In this work, we are interested in studying the spatial structure of current helicity maps. The \textit{z}-related part of current helicity is obtained as a result of several mathematical operations applied to the magnetic field vector data. In addition to the magnetic field measurements uncertainties, which are especially high for the transverse magnetic field, the accuracy of the \textit{H\textsubscript{C}} map depends on the magnetic field pre-processing (especially the 180\degr\ disambiguation routine) and on the computational procedures used to calculate \textit{H\textsubscript{C}}. So, the first question is how reliable the derived current helicity map is? To address this question, we made a comparison of current helicity maps derived from different instruments.

We compared the \textit{H\textsubscript{C}} maps from two space-borne instruments, namely SDO/HMI and Hinode/SOT-SP. Vector magnetograms of AR NOAA 11158 on 2011 February 13 were chosen for calculations of \textit{H\textsubscript{C}} (Fig.~\ref{fig1}). The square contour of the integral \eqref{eq7} was set to 5$\times$5 pixels for SDO/HMI data and to 9$\times$9 pixels for Hinode/SOT-SP data, implying equal sampling contours of 2.5\arcsec$\times$2.5\arcsec on the solar surface. The Hinode/SOT-SP \textit{H\textsubscript{C}} map was rescaled to the SDO/HMI pixel size by an IDL CONGRID function. Then the two maps were co-aligned and cropped to achieve the same field-of-view. At the last step of data reduction, the \textit{H\textsubscript{C}} maps were smoothed with a Gaussian kernel with a width of 2.5\arcsec.

The current helicity maps from SDO/HMI and Hinode/SOT-SP and their comparison are shown in Fig.~\ref{fig1}. One can see an excellent visual agreement between the images, the Pearson's correlation coefficient between the maps is 0.86.

\begin{figure*}
	\includegraphics[width = \linewidth]{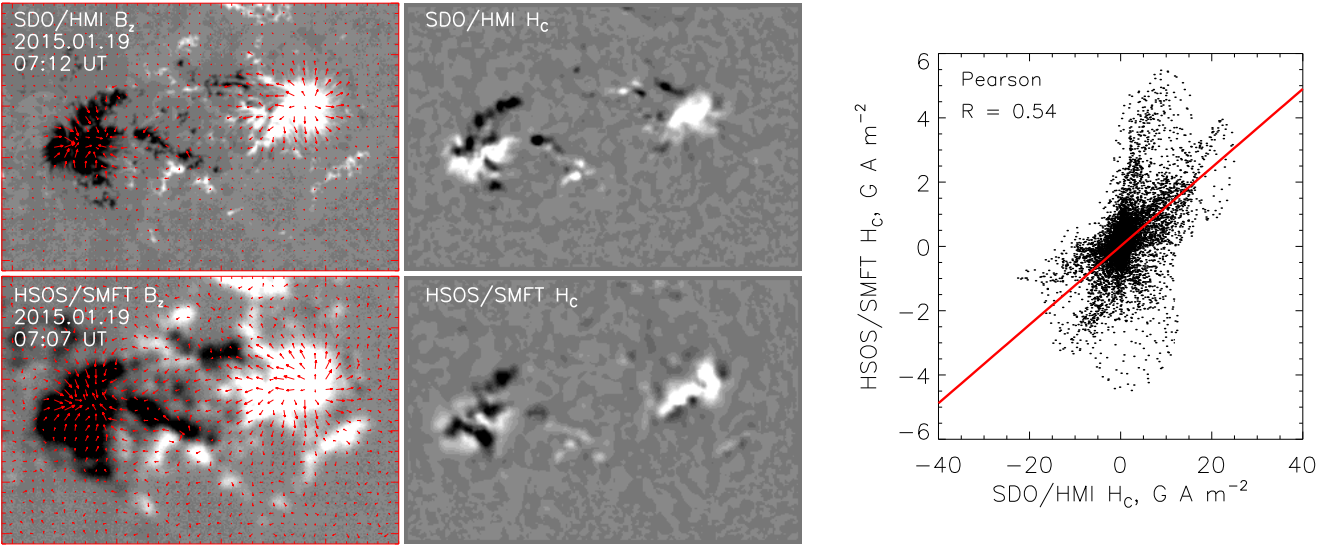}
	\caption{Left: Longitudinal magnetograms of AR NOAA 12266 acquired by SDO/HMI and HSOS/SMFT. The magnetograms were taken on 2015 January 19. The FOV is 150\arcsec$\times$100\arcsec. The magnetograms are scaled from -1000 Mx cm\textsuperscript{-2} (black) to 1000 Mx cm\textsuperscript{-2} for SDO/HMI and from -200 Mx cm\textsuperscript{-2} to 200 Mx cm\textsuperscript{-2} for HSOS/SMFT. Red arrows indicates the direction of the transversal magnetic field. Middle: Current helicity maps of the AR derived from the magnetograms by equation \eqref{eq5}. The maps are scaled from -10 G A m\textsuperscript{-2} (black) to 10 G A m\textsuperscript{-2} for SDO/HMI and from -2.5 G A m\textsuperscript{-2} (black) to 2.5 G A m\textsuperscript{-2} for HSOS/SMFT. Right: A comparison of the current helicity maps. Thick red line shows the best linear fit to the distribution.}
	\label{fig2}
\end{figure*}

We also compared the \textit{H\textsubscript{C}} maps derived from SDO/HMI and from the ground-based HSOS/SMFT instrument. We chose a simple bipolar AR NOAA 12266 observed on 2015 January 19 (Fig.~\ref{fig2}). In this case, the square contour was set to 9$\times$9 pixels for SDO/HMI magnetograms and to 15$\times$15 pixels for HSOS/SMFT data, resulting in contour side length of 4.5\arcsec. This value is consistent with the spatial resolution of the ground-based HSOS/SMFT \citep{Bai2014}. Before rescaling to the pixel size of SDO/HMI, the HSOS/SMFT \textit{H\textsubscript{C}} map was rotated by approximately 6.5\degr\ by an IDL ROT function in order to coalign the $y$-axis of the HSOS/SMFT \textit{H\textsubscript{C}} map and the rotation axis of the Sun. This angle was derived from the SUNPARAM procedure (a part of Yohkoh SolarSoft packages) for the date of observation. Finally, the \textit{H\textsubscript{C}} images were smoothed with 3.5\arcsec\ width Gaussian kernel. The results are shown in Fig.~\ref{fig2}. Although the spatial structure of the \textit{H\textsubscript{C}} maps are similar, the visual agreement is worse than in the previous case. The Pearson's R equals 0.54 suggesting moderate correlation between the data.

Compared to ground-based instruments, space-borne instruments have usually higher resolution and they are not affected by seeing. These factors probably cause different results in the vector magnetic field measurements. This difference can also be found in the calculated distribution of current helicity \citep[see also][]{Xu2016}. There is no way to completely eliminate this discrepancy by data reduction. The comparison of space-borne and ground-based instruments is worth studying in depth and we plan to carry out such an analysis in forthcoming works.

We emphasize that all the instruments use different technique to measure magnetic field and different inversion procedures. Nevertheless, the current helicity maps are quite similar, especially for the space-borne instruments. Moderate agreement between the SDO/HMI and HSOS/SMFT \textit{H\textsubscript{C}} maps can be explained by high uncertainties of the transversal magnetic field measurements in the HSOS/SMFT data due to seeing and, possibly, some systematic errors \citep{Otsuji2015}. Therefore, we can conclude that the obtained \textit{H\textsubscript{C}} maps represent real spatial distribution of current helicity over active regions and the influence of measurement noise or instrumental data contamination is not significant \citep[cf.][]{Pevtsov1994}. Therefore, we can use the obtained maps to analyse the multifractal structures of current helicity.

\section{Results}

We analysed the flatness functions and the variations of flatness exponent of current helicity maps of three ARs, namely NOAA 11158, 12494, 12673. In this part of our study, we used the vector magnetograms from SDO/HMI only. The ARs were tracked during 5--6 days when they were located less than $\pm$30\degr\ from the central meridian. By this reason, we did not apply any deprojection procedures to the magnetic field data. We used the definitive hmi.sharp720s series \citep{Hoeksema2014, Bobra2014}, containing the data on magnetic field strength, inclination, and azimuth. The transversal and longitudinal magnetic fields were calculated under assumption that the $z$-axis coincides with the line-of-sight. The integration contour in equation \eqref{eq7} was set to 5$\times$5 pixels corresponding to 2.5\arcsec\ or approximately 2 Mm contour side length. Before the flatness function calculation, current helicity maps were smoothed with a 5-pixel width Gaussian kernel.

\begin{figure*}
	\includegraphics[width = \linewidth]{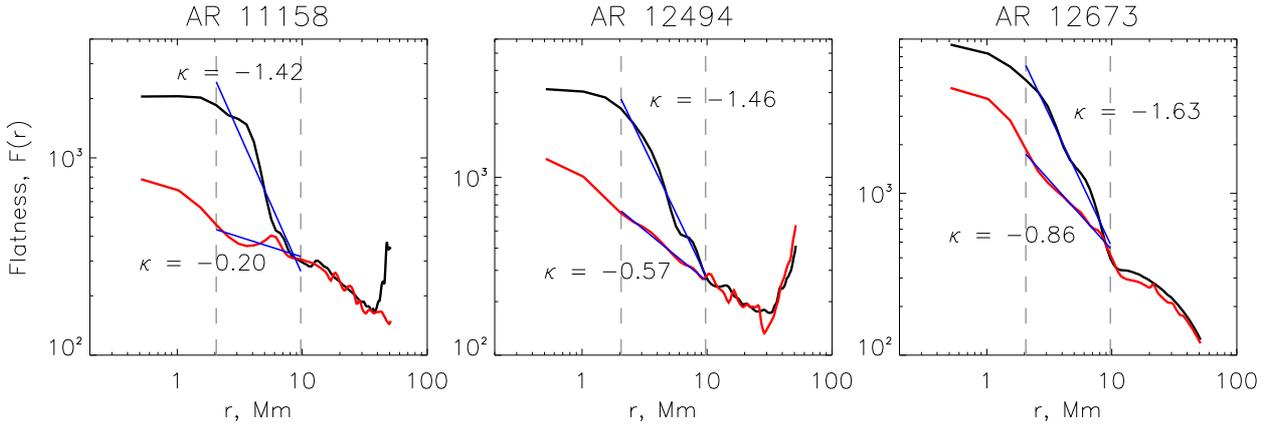}
	\caption{Typical flatness functions (black and red curves) of the current helicity maps for ARs NOAA 11158 (left), 12494 (middle), and 12673 (right). Black curves show flatness functions during the observed dip in $\kappa_{H_C}$, while red curves show typical flatness functions outside the dip (see text and panel b in Figs.~\ref{fig4},\ref{fig5},\ref{fig6}). Blue lines show the best linear fit to the curves within the scale range 2--10 Mm (indicated by grey dashed lines). The flatness functions were calculated for the time moments indicated by dashed grey lines in panels b of Figs.~\ref{fig4},\ref{fig5}, and \ref{fig6}.}
	\label{fig3}
\end{figure*}

We found that the linear size of the current helicity structures rarely exceeded 10 Mm that is in a good agreement with the results by \cite{Pevtsov1994} regarding the size of the current helicity elements size. Structures with a linear size less than 2 Mm are smaller than the integration contour and are not reliable for the analysis. That is why the flatness exponent $\kappa\textsubscript{H\textsubscript{C}}$ was calculated as a linear fit to the flatness function of the current helicity map within the scale range 2--10 Mm (Fig.~\ref{fig3}).

Typical flatness functions derived from the current helicity maps of all three ARs considered in this work are shown in Fig.~\ref{fig3}. One can see that the curves have almost the same slope in the scale range 10--40 Mm implying similar structural complexity at these scales. On the contrary, the slopes in the scale range 2--10 Mm differ significantly suggesting quite different intermittency of small scale current helicity elements at different moments. Saturation of the flatness functions below 2 Mm is caused by artificial smoothing of the current helicity maps at low scales \citep[cf. fig.~8 in ][]{Abramenko2010}.

In addition to the current helicity flatness exponent, we analysed variations of i) the total unsigned flux of an AR, ii) the flatness exponents of longitudinal and transversal magnetic fields, iii) net current helicity of an AR. The total unsigned flux data were retrieved from the headers of the magnetogram FITS files \citep{Bobra2014}. The flatness exponents of the longitudinal ($\kappa\textsubscript{B\textsubscript{z}}$) and the modulus of the transversal ($\kappa\textsubscript{B\textsubscript{t}}$) magnetic fields were determined as a linear fit to the flatness function calculated from corresponding magnetic field maps within the scale range of 5--40 Mm. In the most of the cases, the flatness function of the magnetic field saturated at 5 Mm \citep[see fig.~5 in][]{Abramenko2010}. The effect is caused by relatively high noise of the vector magnetograms \citep{Abramenko2010}. The upper boundary of 40 Mm corresponds to a linear size of the largest spots. Net current helicity was calculated as a sum of current helicity in pixels over the entire map.

\subsection{Active region NOAA 11158}
\begin{figure}
	\includegraphics[width = \columnwidth]{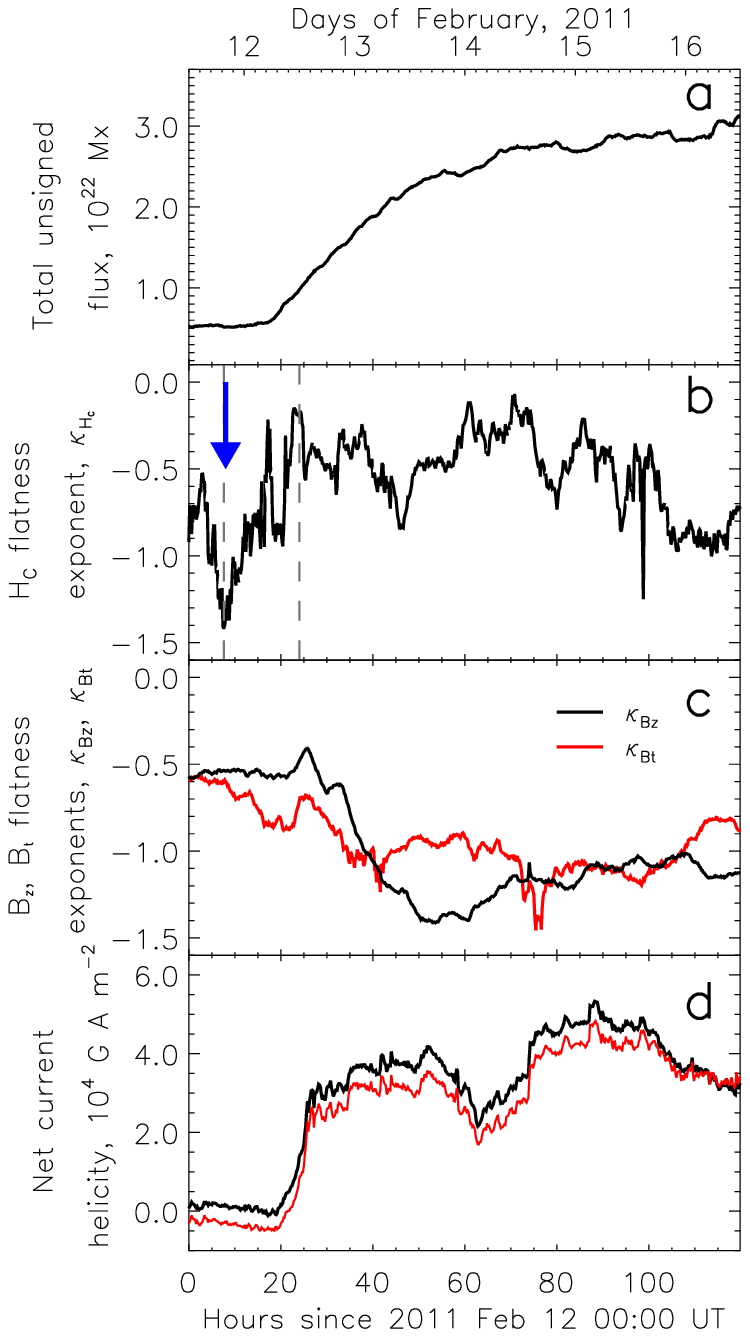}
	\caption{Variations of the total unsigned flux (a), of the current helicity flatness exponent (b), of longitudinal and transversal magnetic field flatness exponents (c), and of net current helicity (d) of AR NOAA 11158. A thick blue arrow (panel b) points the dip in the current helicity flatness exponent curve that preceded the emergence of a new flux. The current helicity flatness functions plotted in Fig.~\ref{fig3}a were obtained at the moments denoted by vertical dashed lines in panel b. Rescaled and shifted net current helicity retrieved from the SHARP data is overplotted in a red color in panel d.}
	\label{fig4}
\end{figure}

AR NOAA 11158 started to emerge in the southern hemisphere on 2011 February 10 as two simple magnetic dipoles that gradually formed a quadrupole magnetic structure. By the beginning of 2011 February 12, the total magnetic flux of the AR reached the value of 5$\times$10\textsuperscript{21} Mx and stayed unchanged until 2011 February 12 17:00 UT. After that, a new intensive emergence started and continued for at least two days (Fig~\ref{fig4}a). The total magnetic flux of the AR reached almost 3$\times$10\textsuperscript{22} Mx by the end of 2011 February 15.

We tracked the AR NOAA 11158 during 5 days from 2011 February 12 to February 16. In total, 600 sets of vector magnetograms were used in the analysis. Variations of the flatness exponent of current helicity $\kappa\textsubscript{H\textsubscript{C}}$ are shown in Fig.~\ref{fig4}b. One can see an abrupt dip of the flatness exponent down to values of -1.4 approximately 12 hours before the start of a new magnetic flux emergence on 2011 February 12 at 17:00 UT. Just before the emergence, the flatness exponent of current helicity restored its initial values and varied in the range from -0.1 to -1.0 up to the end of the observations. The current helicity flatness functions plotted in Fig.~\ref{fig3}a were obtained at the moments denoted by vertical dashed lines in Fig.~\ref{fig4}b.

Note that both flatness function curves in Fig.~\ref{fig3}a have almost the same slope within the scale range above 10 Mm. We found that this slope remains quasi-constant. This implies that, within this range, the intermittency of current helicity remains unchanged.

Flatness exponents of longitudinal, $\kappa\textsubscript{B\textsubscript{z}}$, and of modulus of transversal, $\kappa\textsubscript{B\textsubscript{t}}$, magnetic fields are also shown in Fig.~\ref{fig4}c. Interestingly, before the onset of new emergence both curves show low absolute values of the exponent, about 0.5, and vary slightly. Unlike the $\kappa\textsubscript{H\textsubscript{C}}$ behaviour, a significant increase of $\lvert\kappa\textsubscript{B\textsubscript{z}}\rvert$ and $\lvert\kappa\textsubscript{B\textsubscript{t}}\rvert$ is observed synchronously with the increase of the total magnetic flux. Note also that the increase of $\lvert\kappa\textsubscript{B\textsubscript{z}}\rvert$ precedes M- and X-class flare events on 2011 February 14 17:28 UT (M6.6) and on 2011 February 15 01:44 UT (X2.2). This is consistent with the results by \cite{Abramenko2010}: an increase of absolute $\kappa\textsubscript{B\textsubscript{z}}$ in ARs is often followed by strong flare events.

The net current helicity variations of the AR calculated using equation \eqref{eq7} are plotted in Fig.~\ref{fig4}d. Before the onset of a new flux emergence, the net current helicity fluctuated around zero values, implying the helicity balance. After the emergence onset, the helicity balance changed rapidly and positive helicity started to prevail obeying the hemispheric sign rule for current helicity. The net current helicity retrieved from the headers of FITS file is overplotted in panel d of Fig.~\ref{fig4} in red colour. The curve was rescaled and artificially shifted down to visualize a good agreement between the net current helicity obtained by differential and integral formulae.

\subsection{Active region NOAA 12494}

\begin{figure}
	\includegraphics[width = \columnwidth]{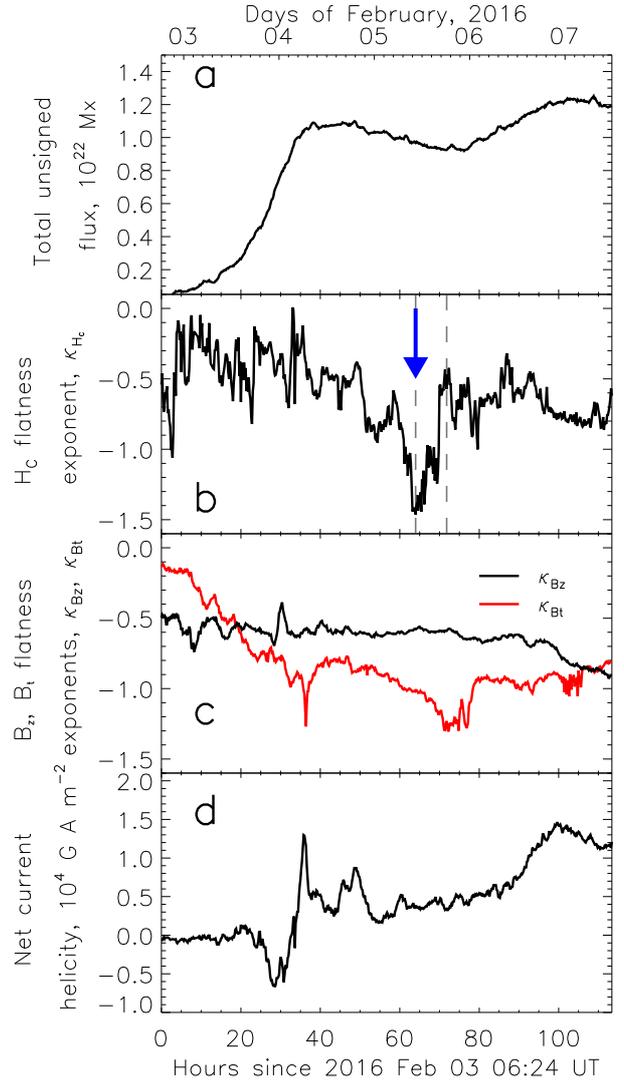}
	\caption{The same as in Fig.~\ref{fig4} for AR NOAA 12494.}
	\label{fig5}
\end{figure}

An AR NOAA 12494 started to emerge on 2016 February 03 and exhibited the total magnetic flux of approximately 1.1$\times$10\textsuperscript{22} Mx by the beginning of 2016 February 05 (Fig.~\ref{fig5}a). The AR had the well-formed leading and following spots that made up a simple magnetic dipole and a pore-like mixed-polarity magnetic structures between the main sunspots. During the development of the AR, the leading spot was more fragmented than the following one. The AR started to decay at about 00:00 UT on 2016 February 05. The decay was interrupted by a new growth of the total unsigned magnetic flux. Visual inspection of the AR magnetograms reveals an emergence of a new magnetic dipole in a close vicinity of the following spot. This event probably caused a quick elongation and fragmentation of the following spot.

Similar to the previous case, a rapid increase of the modulus of the current helicity flatness exponent was observed approximately 12 hours before the onset of a new flux emergence (Fig.~\ref{fig5}b, a blue arrow): $\kappa\textsubscript{H\textsubscript{C}}$ reached the value of -1.4 for a short time interval and then restored its initial value of about -0.5. Note that the flatness exponent of the longitudinal magnetic field remains low during the entire observational interval. The flatness exponent of the modulus of the transversal magnetic fields neither reflects the pre-emergence variations of $\kappa\textsubscript{H\textsubscript{C}}$. This AR, being located at the southern hemisphere, also obeyed the hemispheric sign rule for current helicity.

\subsection{Active region NOAA 12673}

\begin{figure}
	\includegraphics[width = \columnwidth]{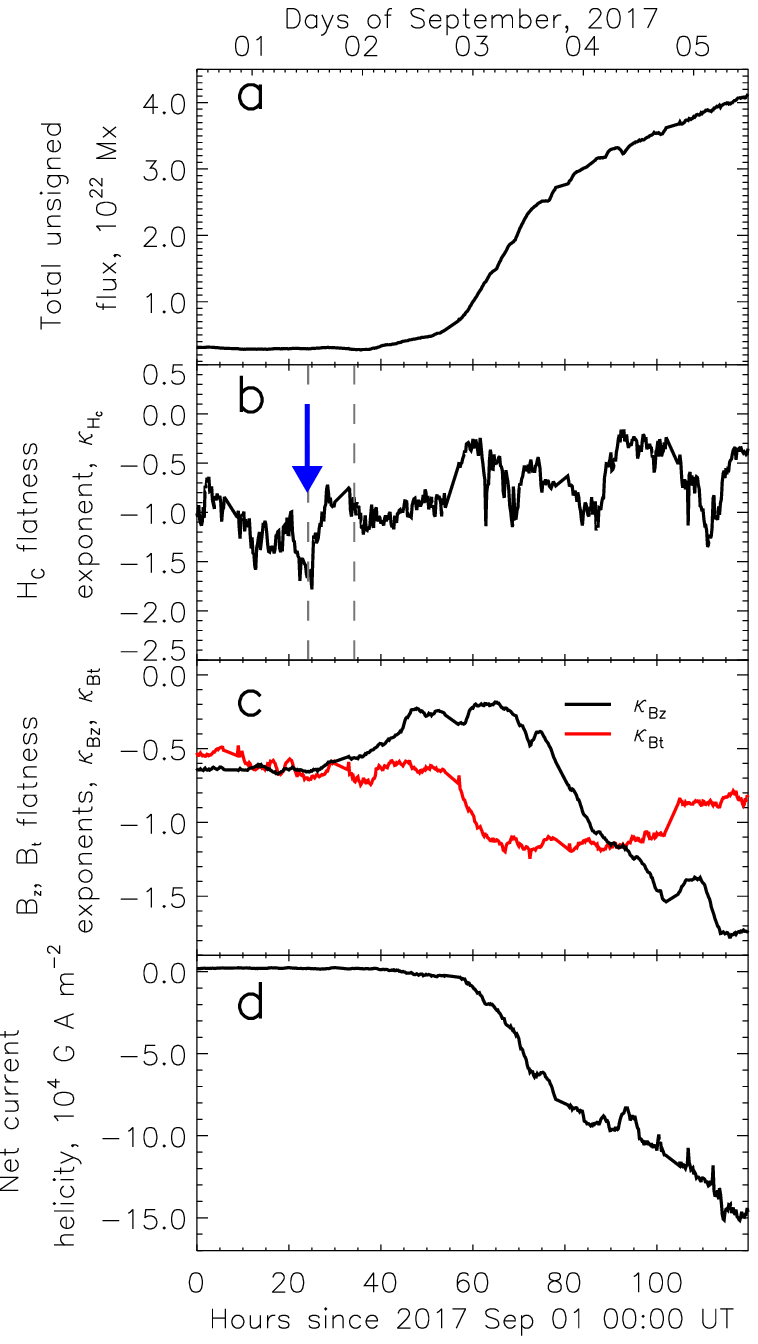}
	\caption{The same as in Fig.~\ref{fig4} for AR NOAA 12673.}
	\label{fig6}
\end{figure}

AR NOAA 12673 was one of the most interesting AR of cycle 24. The AR appeared at the eastern limb as a simple unipolar decaying positive-polarity spot of approximately 3$\times$10\textsuperscript{21} Mx on 2017 August 29. We tracked the AR for 5 days since 2017 September 01 00:00 UT. As the AR moved westward, a new magnetic flux started to emerge southeast of the spot on 2017 September 02. The flux emergence rate was greater than any value ever reported so far \citep{Sun2017}. Two bipolar emerging regions formed a complex quasi-circular topological system \citep{Yang2017}. The AR produced a series of X-class flares one of which (X9.3 SOL2017-09-06T11:53) was the strongest event since 2005 \citep{Sun2017}. One can see in Fig.~\ref{fig6}c that the AR displayed the significant increase of the intermittency of longitudinal magnetic field prior to the first strong flare event. By the end of 2017 September 07, the total unsigned flux of the AR increased up to 6$\times$10\textsuperscript{22} Mx.

Although the flatness exponent of current helicity had on average higher value than in previous two cases, the same tendency is observed: the $\kappa\textsubscript{H\textsubscript{C}}$ curve exhibits an abrupt dip 15--20 hours before the start of new emergence (Fig.~\ref{fig6}a,~b). Again, neither flatness exponents of magnetic field components nor net helicity display any significant changes at the moment when the dip occurs (Fig.~\ref{fig6}b, c).

Interestingly, although the AR emerged at the southern hemisphere, its net current helicity was negative, i.e. the AR violated the current helicity hemispheric sign rule.

\section{Conclusions and discussion}

In this article, we analysed intermittency of the $z$-related part of current helicity in solar active regions. First of all, we made a comparison of current helicity maps derived from different instruments. We concluded that the maps predominantly represent persistent pattern of current helicity rather than random structures.

We used the high-order structure functions approach to obtain the intermittency (or multifractality) spectra -- the flatness functions -- of current helicity maps. The variations of the slope of flatness functions -- the flatness exponent $\kappa_{H_C}$ determined for the scale range of 2--10 Mm -- were scrutinized. This parameter is the measure of intermittency of current helicity. In addition, we analysed the co-temporary variations of the flatness exponents of longitudinal and of the modulus of transversal magnetic fields as well as changes in the net current helicity and in the total unsigned magnetic flux of ARs.

Three ARs NOAA 11158, 12494, and 12673 were studied. Two of them, namely NOAA 11158 and 12494, obeyed the hemispheric rule for net current helicity sign while AR NOAA 12673 violated the rule. It is worth noting that two ARs of high-flaring activity (NOAA 11158 and 12673) exhibited rapid increase of intermittency of longitudinal magnetic field prior to a series of strong flares, while  intermittency of longitudinal magnetic field of the low-flaring AR NOAA 12494 remained at a relatively low level. This finding is in a general agreement with the results by \cite{Abramenko2010}.

All the ARs had a common feature in their evolutionary behaviour: during the observational interval, emergence of a new magnetic flux has been observed in a close vicinity of the pre-existing magnetic structures. In all three cases, we observed an abrupt dip in the current helicity flatness exponent $\kappa_{H_C}$ time profile approximately 12 hours prior to the emergence onset. This result resembles the findings by \citet{Singh2016}, who analysed the distribution of line-of-sight magnetic field within patches that enclosed emerging AR. They found that the kurtosis, i.e. the fourth statistical moment, of magnetic field increases prior to the formation of AR. In our opinion, the effect is caused by small-scale strong magnetic features that appear at the photosphere during the first stages of AR emergence \citep[e.g.][]{Zwaan1985, Lites1998}. These features might make the magnetic field distribution more heavy-tailed leading to the kurtosis increase. In this work, we also did find changes in the ratio of the statistical moments of current helicity -- a quantity derived from the magnetic field. However, unlike to results by \citet{Singh2016}, we found pre-emergence changes in the structure of electric currents rather than in its value. Note also that these structural changes take place before any signatures of new emerging flux appear at the solar surface, i.e. they might reflect the subphotospheric dynamics of magnetic fields.

The observed decrease in the flatness exponent value from, say, -0.5 to -1.5, means the steepening of the intermittency spectrum, and, therefore, the increase of complexity \citep{Abramenko2005}. We may propose the following qualitative explanation of this phenomenon. Current helicity is a product of electric current density and magnetic field. In absence of any disturbances, magnetic fields exhibit gradual long-term evolution of their spatial structure while electric currents are changing more rapidly. From the formal point of view, a current helicity map represents the electric current distribution weighted by the magnetic field magnitude (here we do not take into account the sign of the magnetic field). In other words, the current helicity structure replicates the fast-changing structure of electric currents inside regions of quasi-uniform magnetic field.

The increase of the modulus of the $H\textsubscript{C}$ flatness function slope within some scale range implies an increase of the complexity and intermittency of current helicity at these scales. We suppose that the observed dip in the current helicity flatness exponent curve can be explained by changing of topology of electric currents, namely by fragmentation and realignment of current structures. This fragmentation might be caused by disturbances produced by emerging magnetic flux as it lifts through the convection zone. The simplest physical explanation for such disturbances might be an excitation of near-surface inductive currents as the subphotospheric magnetic setup changes. More generally, we suppose that the changes of electric current system precede the flux emergence.

If our reasoning is correct, we can make two important conclusions. First, the vertical electric currents observed at the photosphere level extend into the convection zone. Second, the rapid changes in the distribution of electric currents can be used as a tool for prediction of a new magnetic flux emergence.

\section*{Acknowledgements}

We thank Alexei Pevtsov for helpful discussions and suggestions. We also thank anonymous referee for his/her helpful comments. SDO is a mission for NASA's Living With a Star (LWS) program. The SDO/HMI data were provided by the Joint Science Operation Center (JSOC). \textit{Hinode} is a Japanese mission developed and launched by ISAS/JAXA, with NAOJ as domestic partner and NASA and STFC (UK) as international partners. It is operated by these agencies in co-operation with ESA and NSC (Norway). This study was supported in part by the Russian Fund for Basic Research (RFBR) projects 17-52-53203 and 17-02-00049, by the National Natural Science Foundation of China projects 11673033, 11703042, 11427901, U1731241, by the cooperative NSFC-RFBR grant 11611530679 and by the Strategic Priority Research Program of the Chinese Academy of Sciences XDA15052200. A.K. and K.K. would like to thank the staff at Huairou Solar Observing Station of National Astronomical Observatories of China for warm hospitality and excellent working environment during their visit.









\bsp	
\label{lastpage}
\end{document}